\documentclass[10pt]{article}

\usepackage{graphicx}
\usepackage{authblk}
\usepackage{multirow}
\usepackage[fleqn]{amsmath}
\usepackage[margin=3.5cm]{geometry}
\usepackage{natbib}
\usepackage{lineno}
\usepackage{float}

\def \hat {\widehat}
\def \bbeta{\mbox{\boldmath$\beta$}}

\def \bvarepsilon{\mbox{\boldmath$\varepsilon$}}
\def \bzeta{\mbox{\boldmath$\zeta$}}
\def \bxi{\mbox{\boldmath$\xi$}}
\def \bepsilon{\mbox{\boldmath$\epsilon$}}

\def \blambda{\mbox{\boldmath$\lambda$}}

\def \bdelta{\mbox{\boldmath$\delta$}}
\def \balpha{\mbox{\boldmath$\alpha$}}

\def \bmu{\mbox{\boldmath$\mu$}}
\def \bsigma{\mbox{\boldmath$\sigma$}}

\def \btau{\mbox{\boldmath$\tau$}}
\def \bdelta{\mbox{\boldmath$\delta$}}
\def \btau{\mbox{\boldmath$\tau$}}

\def \bvarepsilon{\mbox{\boldmath$\varepsilon$}}
\def \bxi{\mbox{\boldmath$\xi$}}

\def \bv{\mathbf{v}}

\def \bY{\mathbf{Y}}
\def \bX{\mathbf{X}}
\def \bZ{\mathbf{Z}}

\def \bI{\mathbf{I}}

\def \b1{\mathbf{1}}
\def \bY{\mathbf{Y}}

\title{\large{\textbf{Analysis of longitudinal multilevel experiments using GAMLSSs}}  }

\author[1]{{\normalsize{Gustavo Thomas}}\thanks{Corresponding author. E-mail: gustavothomas17@usp.br}}
\author[2]{{\normalsize{Alexandre Igor de Azevedo Pereira}}}
\author[1]{{\normalsize{Cristian Marcelo Villegas Lobos}}}
\author[1]{{\normalsize{Clarice G.B. Dem\'{e}trio}}}
\affil[1]{{\small Department of Exact Sciences, ESALQ/USP, Piracicaba, SP, Brazil}}
\affil[2]{{\small Department of Agronomy, IF Goiano, Uruta\'{i}, GO, Brazil}}

\date{}

\begin{document}
	
\maketitle

\section*{Abstract}

The standard procedures for analysing hierarquical or grouped data are by (non)linear mixed models or generalized mixed models. However, the generalized additive models for location, scale and shape (GAMLSSs) also allow different types of random effects to be included in the model formulation. Even though already popular in many areas of research, this type of models have not been found to be used for mixed modeling purposes yet. Therefore, this paper describes the analysis of an experiment with plants' growth using mixed GAMLSSs, comparing it to a linear mixed model approach.

\hspace{-1.5cm}\noindent \textbf{Keywords:} linear mixed models; mixed GAMLSSs; grouped data; sweet corn growth
\hspace{-1.5cm}

\section{Introduction}

Statistical analysis of experiments using linear mixed effects models date back at least to \citet{Jackson39}, who defined a model with one factor as having fixed effects and another as a sample from a normally distributed random variable. Since then, mixed modeling has become a major area of statistical research, including work on computation of maximum likelihood estimates, non-linear mixed effect models, missing data in mixed effects models, and Bayesian estimation of mixed models. These models are applied in many disciplines where multiple correlated measurements are made on each unit of interest. 
A brief history of linear mixed models can be found in \citet{West2014}.

Mixed models for non normal data were more recently developed, though. \citet{Schall91} was one of the first to present the estimation of generalized linear models with random effects. Usually referred to as generalized linear mixed models (GLMMs), this type of regression models were popularized by the works of \citet{Breslow93}, \citet{McCulloch2001} and \citet{Bolker2009}, to name a few. 

Linear, nonlinear and generalized mixed models are currently the standard approaches for analysing experiments with grouped data. \citet{Rees99}, \citet{Buckley03}, \citet{Schlenker09}, \citet{Strathe10} and \citet{Paine12} are some examples. These types of models restrict the number of possible distributions to be assumed for the response variable to the ones belonging to the exponential family of distributions. Furthermore, they only allow the mean parameter of the response distribution to be modeled explicitly with functions of the available covariates.

The generalized additive models for location, scale and shape (GAMLSSs) introduced by \citet{Rigby05} allow any computable probability distribution to be assumed for the response variable and model all its parameters (up to four) with a separate predictor for each. In fact, there are more than 90 distributions available in the current implementation of the \verb|gamlss.dist| package in \verb|R| and anyone can create a new one. While GAMLSSs are already well established in the literature for centile growth curve estimation (frequently used with this purpose by the WHO - World Health Organization) and have been gaining popularity in other areas as well (such as industrial \citep{Barajas15}, medical \citep{Petterle16}, financial \citep{Gilchrist09}, foresty \citep{Hudson09}, etc), no applications where GAMLSSs were used for mixed modeling purposes have been found. Since the \verb|gamlss::gamlss()| (\verb|R package::function within package|) function allows an interface to be made with the well known \verb|nlme::lme()| function, it can be used for repeated measurements, multilevel modeling, random intercept and slopes, etc, among other mixed model fit purposes. The present work aims to ilustrate how this GAMLSSs' methodology can be used to analyze a multilevel sweet corn experiment with repeated measures comparing this analysis to one where a classical mixed model approach was employed.

\section{Experiment description}

The sweet corn experiment analyzed was carried out from March to May of 2015 in greenhouses in the Instituto Federal Goiano (IF Goiano) in Goias State, Brazil. Lack of water situations in sweet corn plants is difficult to monitor due to climatic events such as El Niño \citep{Li2011}, leading to considerable losses. Therefore, this study aimed to test the hypothesis that potassium silicate applications could elicit the resistance of these plants, reducing environmental stress effects caused by the lack of water. The experimental design consisted of a split-plot with subsampling. The area inside the greenhouse was divided in four blocks, four plots within each block, four subplots within each plot and fourteen sweet corn plants were planted in each subplot. The sweet corn plants represented the subsamples in each subplot. Four levels (15, 30, 45 and 60 kPa) of soil water tensions were randomized to plots and four doses of potassium silicate (0, 6, 12 and 24 L/ha) were randomized to subplots. Four of the fourteen plants in each subplot were randomly selected and their heights were measured (in the same four plants in each subplot at all times) at 30, 45, 60, 75 and 90 days after seeding. The objective of the experiment was to evaluate the relationship between sweet corn plant age and its development under induced water stress and leaf potassium silicate applications. 

\begin{figure}
	\centering
	\includegraphics[width=12cm]{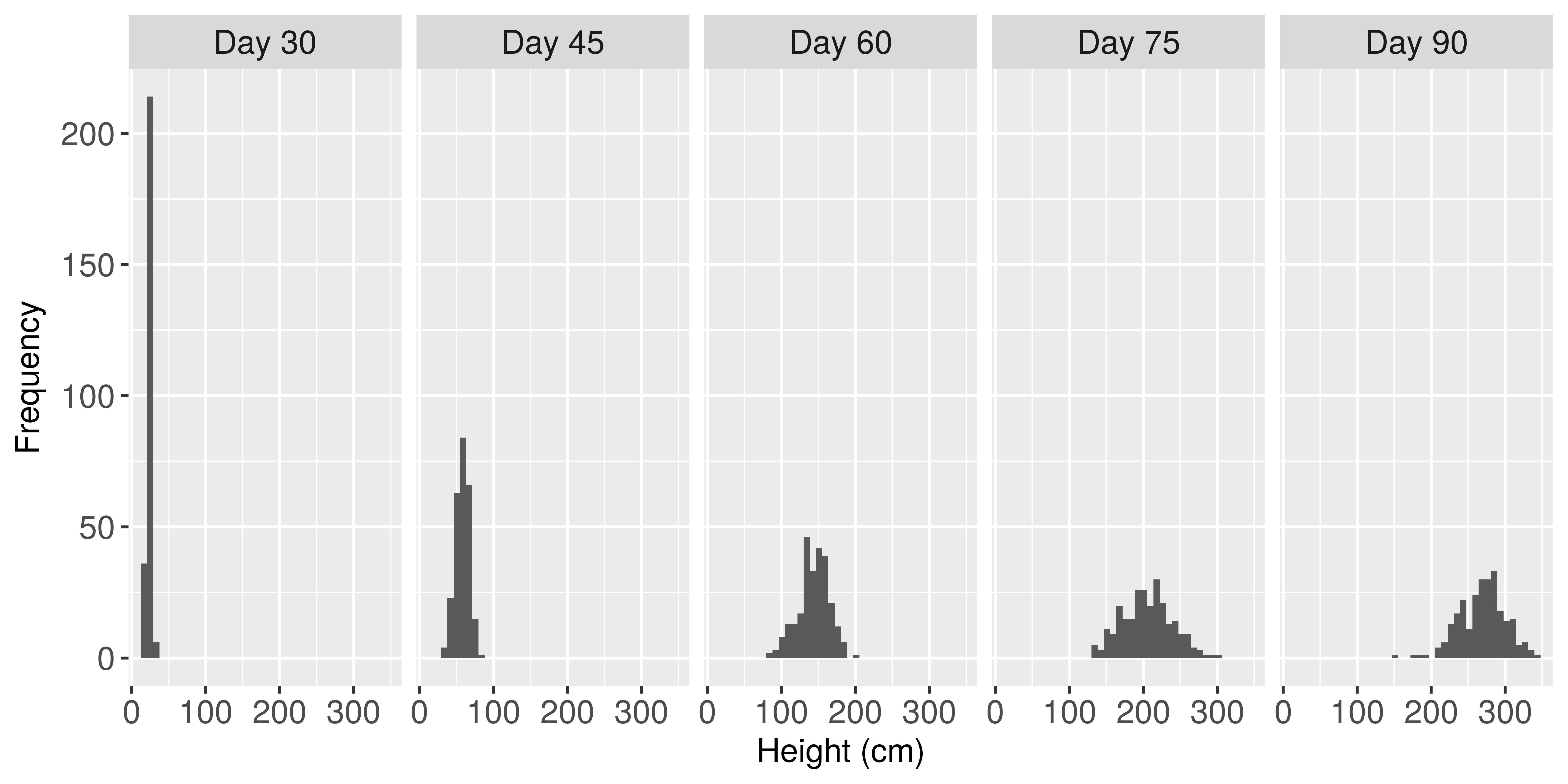}
	\caption{Histograms of height per time.}
	\label{hists.time}
\end{figure}

\begin{figure}
	\centering
	\includegraphics[width = 11cm, height = 9cm]{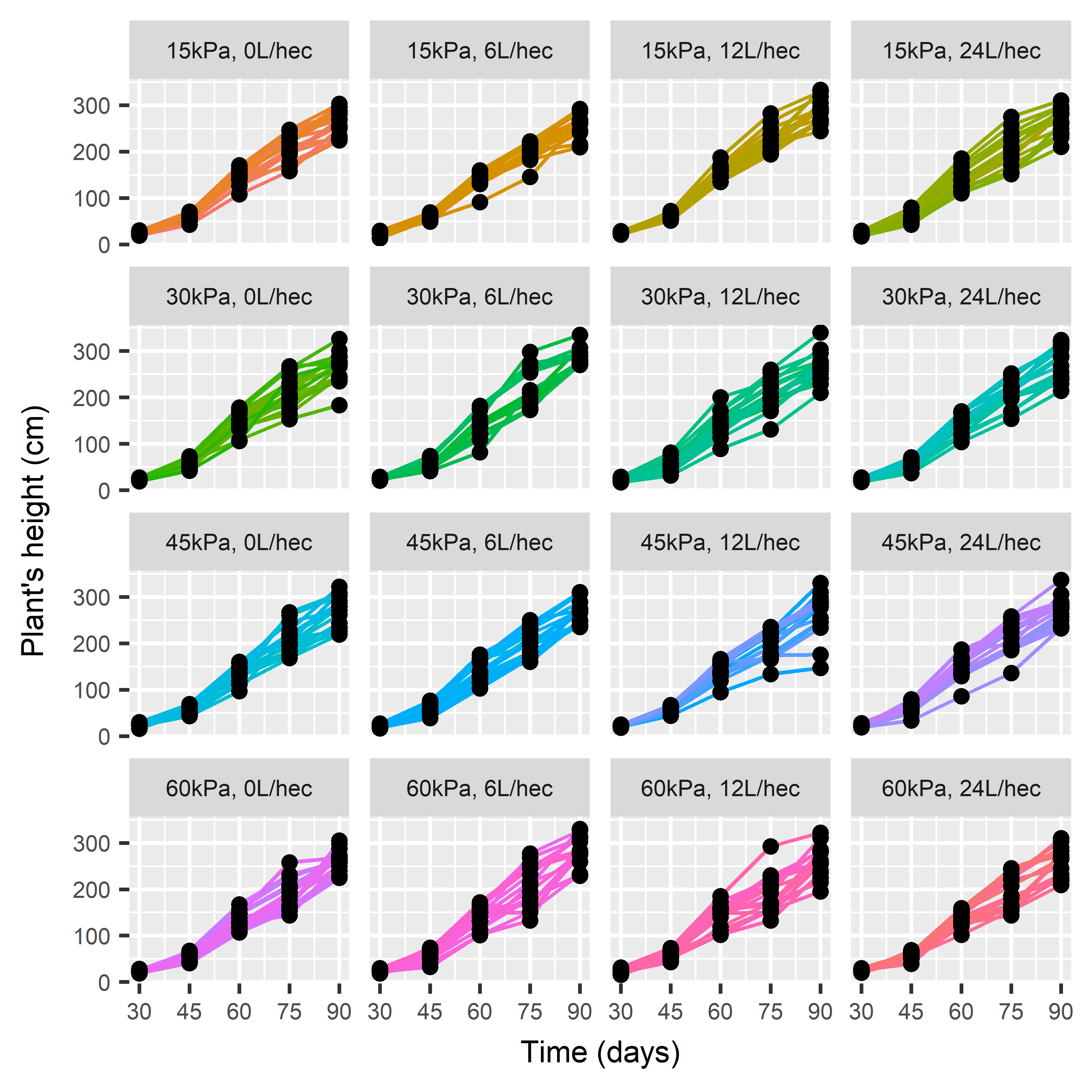}
	\caption{Plant's profile growth per treatment.}
	\label{ht}
\end{figure}

\begin{figure}
	\centering
	\includegraphics[width=11cm,height=8cm]{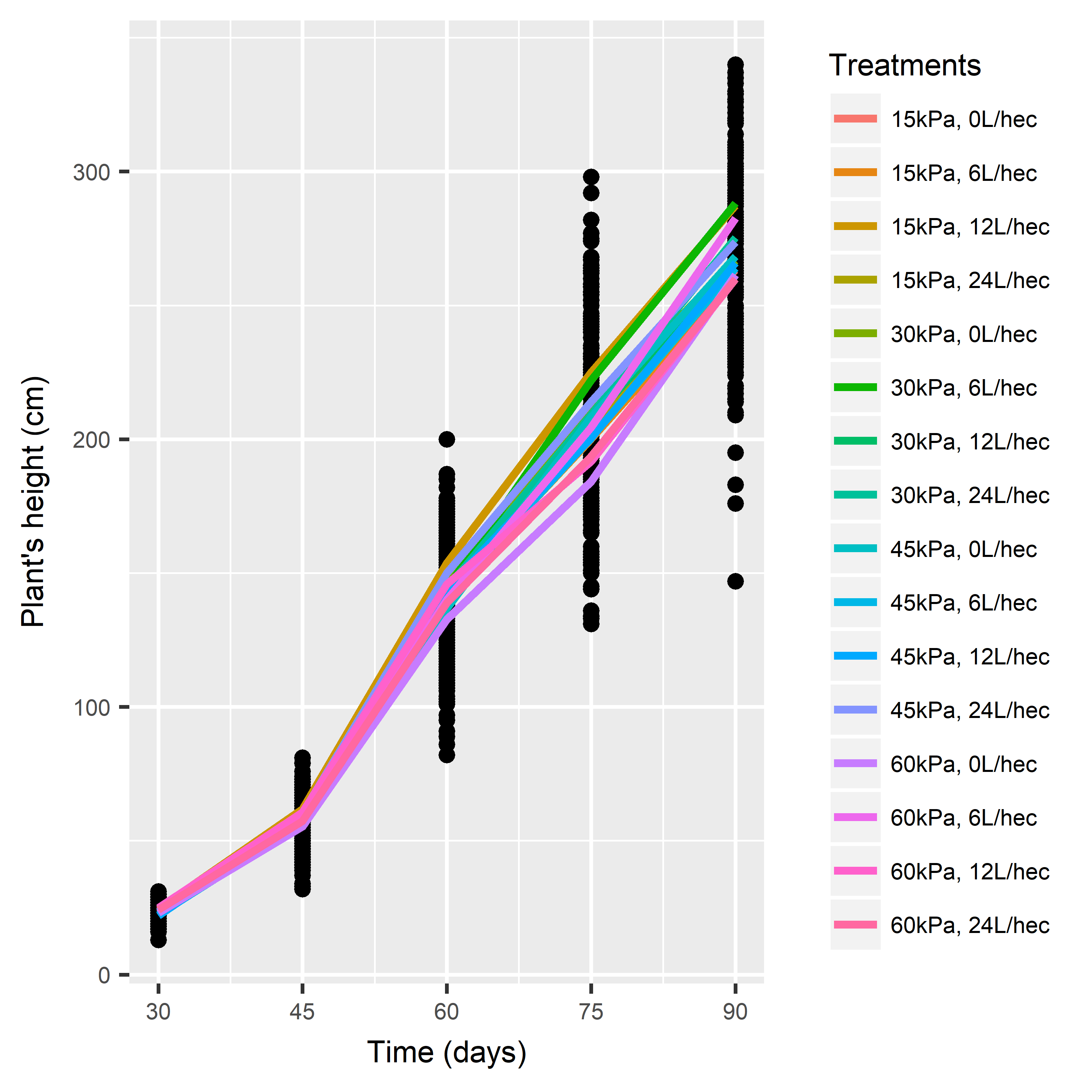}
	\caption{Average growth lines of each treatment combination.}
	\label{height.treat}
\end{figure}

Initially, an exploratory analysis of the data set was conducted. Histograms of the height measurements at each time of observation are displayed in Figure \ref{hists.time}. The histograms highlight that the height distributions are roughly symmetric at all times but with increasing variability. For a visual assessment of treatment effects, growth plots of the 256 plants measured in the experiment were produced and are displayed by treatment (combination of soil tension and silicate dose levels) in Figure \ref{ht}. Each line on each plot of this figure links the 5 height measurements (5 dots) of a plant under a specific treatment (displayed at the top of each plot). There does not seem to be large differences between plant's development across treatments, since on average the plants show similar growth patterns and reach around 300 centimeters of height at 90 days after seeding. Furthermore, the growth pattern reveals that the highest height increase occurs between 45 and 60 days after seeding. This nonlinear pattern can be better seen in Figure \ref{height.treat}, which shows the mean growth profiles for each combination of soil tension and potassium silicate levels (16 treatment lines). In accordance with Figure \ref{ht}, this one shows that there are not strong treatment effects but there is a mild triple interaction effect of both factors and time, since there are some crossings of the mean treatment lines in the right side of the graph.

\section{Analysis}

A common procedure for the analysis of multilevel experiments is to define random effects for all terms justified by the randomization (\citet{Piepho03}, \citet{Brien09}). For a classical split-plot experiment, this would mean one random effect for mainplots and one for subplots, which corresponds to a two-level model in the notation of \citet{Pinheiro00}. Since in this sweet corn experiment there are also subsamples (the sweet corn plants) within each subplot, a 3rd level is characterized. Therefore, the random effects part of the model was formulated with a random effect for main plots, one for subplots and another for subsamples. The fixed part of the model contains the main effects of blocks (controlling a soil fertility gradient), soil tensions, potassium silicate doses, the interaction of the last two and an appropriate function to model the effect of time, possibly interacting with the factors as well. A mixed model for this experiment is given by
\begin{equation}
y_{ijklm}= \mu + \beta_i + \alpha_j + \epsilon_{ij} + \lambda_{k} + \alpha\lambda_{jk} + \varepsilon_{ijk} + \xi_{ijkl} + f(\tau_m) + \zeta_{ijklm}, \\
\label{mod}
\end{equation}
where $y_{ijklm}$ is the $m$-th height measurement of the $l$-th plant located at the $i$-th block, $j$-th plot and $k$-th subplot ($i$, $j$, $k$ and $l$ vary from 1 to 4 and $m$ from 1 to 5); 
$\mu$ is a constant common to all observations; 
$\beta_i$ is the fixed effect of block $i$; 
$\alpha_j$ is the fixed effect of the $j$-th soil tension level; 
$\epsilon_{ij}$ is the error associated to plot $ij$; 
$\lambda_{k}$ is the effect of the $k$-th potassium silicate level; 
$\alpha\lambda_{jk}$ is the interaction of soil tension and potassium silicate at levels $j$ and $k$, respectively; 
$\varepsilon_{ijk}$ is the error associated with the subplot $ijk$; 
$\xi_{ijkl}$ is the error associated with the subsample (plant) $ijkl$; 
$f(\tau_m)$ is a function of time (days when measurements were taken) and 
$\zeta_{ijklm}$ is the `residual error' associated with the $m$-th height measurement on plant $ijkl$.

The first three error terms in Equation (\ref{mod}) are assumed to be random variables with distributions $\epsilon_{ij} \sim \mathcal{N}(0,\sigma^{2}_{P})$, $\varepsilon_{ijk} \sim \mathcal{N}(0,\sigma^{2}_{SP})$ and $\xi_{ijkl} \sim \mathcal{N}(0,\sigma^{2}_{SS})$. They are also supposed to be independent among themselves and independent of the residual errors, which are assumed to follow $\zeta_{ijklm} \sim \mathcal{N}(0,\sigma^{2})$.

In matrix notation, this mixed model is written as
\begin{equation}
\bY= \bX_G\bmu + \bX_B\bbeta + \bX_A\balpha + \bX_L\blambda + \bX_{AL}\balpha\blambda + \bX_{T}\btau + \bZ_{SP}\bvarepsilon + \bZ_P\bepsilon + \bZ_{SS}\bxi + \bzeta, \\
\label{modm}
\end{equation}
where the $\bX$'s and $\bZ$'s correspond to design matrices for the fixed and random effects, respectively. The mean of the response vector is given by E$[\bY]=\bX_G\bmu + \bX_B\bbeta + \bX_A\balpha+ \bX_L\blambda + \bX_{AL}\balpha\blambda + \bX_{T}\btau$ and the variance-covariance matrix by Var$[\bY]=\sigma^2_{P}\bZ_P\bZ_P' + \sigma^2_{SP}\bZ_{SP}\bZ_{SP}' + \sigma^2_{SS}\bZ_{SS}\bZ_{SS}' + \sigma^2\bI_{256}$. The variance-covariance structure resulting from this matrix does not correspond to any known structure, therefore an unstructured or general symmetric positive-definite matrix was initially assumed for Var[$\bY$] in the analysis. 

Even though Figures \ref{ht} and \ref{height.treat} gave a visual idea of the time effect, the number of possible functions to be tested for $f(\tau_m)$ is high. Furthermore, from Figure \ref{height.treat} there is some evidence of an interaction effect between time and the treatment factors. 
In this analysis, linear polynomial functions were used to model the effect of time. Two analyses were performed separately using the \verb|nlme::lme()| and \verb|gamlss::gamlss()| functions. These strategies are described in the following sections.

\subsection{Using function \textit{nlme::lme()}}
\label{lme}

The \verb|nlme::lme()| function is among the most popular to fit linear mixed-effects models in \verb|R|. The function was implemented under the general framework of \citet{Lindstrom1988}, but some of its features were originally documented by different authors. For example, the model formulation was first described in \citet{Laird1982}, the variance-covariance parameterizations were described in \citet{Pinheiro1996} and the use of the variance functions implemented in the \verb|lme()| function were first presented by \citet{Davidian1995}. However, the most complete reference of the function and the \verb|nlme| package as a whole is given by \citet{Pinheiro00}. The authors give an easy-going introduction to linear and nonlinear mixed models through analyses of real datasets using the \verb|nlme| library, explaining the capabilities of the library's functions and theoretical aspects along the way. 

This section describes the analysis of the sweet corn experiment using linear mixed effects models (LMMs) fitted by maximum likelihood using the \verb|nlme::lme()| function. The first mixed model fitted to the sweet corn data differed from (\ref{mod}) only in that it had the triple interaction of time with potassium silicate and soil tension factors and the double interactions of time with each of the factors along with the main effect of time. Following the methodology described in \citet{Pinheiro00} section 2.4, the significance of fixed effects' terms in the model was assessed by conditional F-tests using sequential sum of squares. Table \ref{lmm} shows the F-tests results for the fixed terms in the model in the order they were included. The bottom line of this table shows that the triple interaction tested was highly significant. Since time is a quantitative covariate, this means that the plants' growth trend differed significantly using at least 2 different combinations of soil tension and potassium silicate levels, that is, there are significantly different growth trends in plants that received different treatments. 

\begin{table}[ht!]
	\centering
	\caption{F-tests of fixed effects terms in the classical LMM.}
	\label{lmm}
	\begin{tabular}{lllll}
		\hline
		Term           & Degrees of freedom & F-value    & p-value    \\ \hline
		intercept    & 1      & 5,716.53  & \textless.0001 \\
		block          & 3      & 3.20      & 0.0623         \\
		time          & 1      & 19,669.19 & \textless.0001 \\
		tension           & 3      & 0.52      & 0.6716         \\
		silicate            & 3    & 0.18      & 0.9113         \\
		time:tension     & 3      & 3.05      & 0.0279         \\
		time:silicate      & 3    & 0.88      & 0.4510         \\
		tension:silicate       & 9     & 1.23      & 0.2749         \\
		time:tension:silicate & 9     & 2.56      & 0.0065    \\   \hline
	\end{tabular}
\end{table}

\begin{figure}[!htb]
	\centering
	\includegraphics[width=10cm,height=7cm]{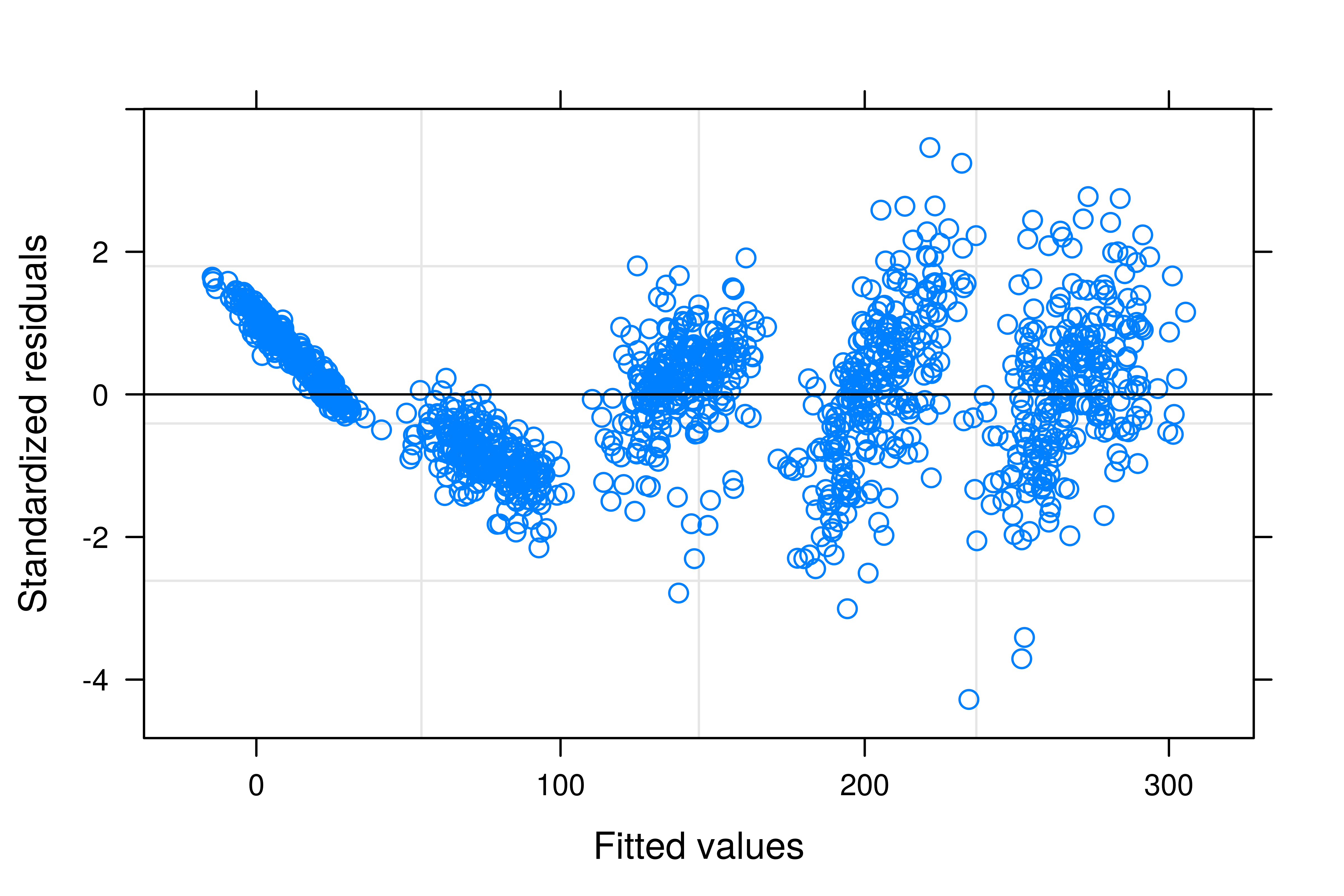}
	\caption{Residuals versus fitted values graph for the classical linear mixed model.}
	\label{res_clmm}
\end{figure}

A graph of the standardized residuals plotted against the fitted values of this model is shown in Figure \ref{res_clmm}. From this plot five groups of residuals are identifiable, corresponding to each moment of height measurement. At 30 days the plant's height varied from 0 to approximately 40cm, at 45 days it varied from around 50 until 100cm and so on until the last measurements from around 220cm onwards. This plot also highlights at least 3 problems with the model fitted. Two of them are the size of the residuals (some of them are below -4) and the heterogeneity of variance among the groups of observations (days of measurement). The third is that the model did a very poor job at predicting plant heights below 100cm, specially the ones taken at 45 days. This happens because this linear mixed model is assuming a constant growth rate during the entire period of 90 days, which is not consistent with Figure \ref{height.treat}. The growth rate between 30 and 45 days was definitely lower than between 45 and 60 days, which caused the model to overestimate the heights at 45 days resulting in the group of negative residuals in Figure \ref{res_clmm}.

To account for the different growth rates of the sweet corn plants, a quadratic and a cubic function of time were added to the fixed part of the model. Table \ref{lmm2} shows the related F-tests' results for these terms along with the triple interaction of the linear effect of time with both factors. It is concluded that both polynomial effects are needed in the model and their inclusion increased the contribution of the triple interaction to the model. 

\begin{table}[ht!]
	\centering
	\caption{F-tests for the triple interaction and polynomial functions of time in the classical LMM.}
	\label{lmm2}
	\begin{tabular}{lllll}
		\hline
		Term           & Degrees of freedom & F-value    & p-value    \\ \hline
		time:tension:silicate  & 9     & 3.750      & 0.0001         \\
		time$^2$       	   & 1     & 69.15      & \textless.0001         \\
		time$^3$ 	   & 1     & 180.53      & \textless.0001    \\   \hline
	\end{tabular}
\end{table}

\begin{figure}[!htb]
	\centering
	\includegraphics[width=10cm,height=7cm]{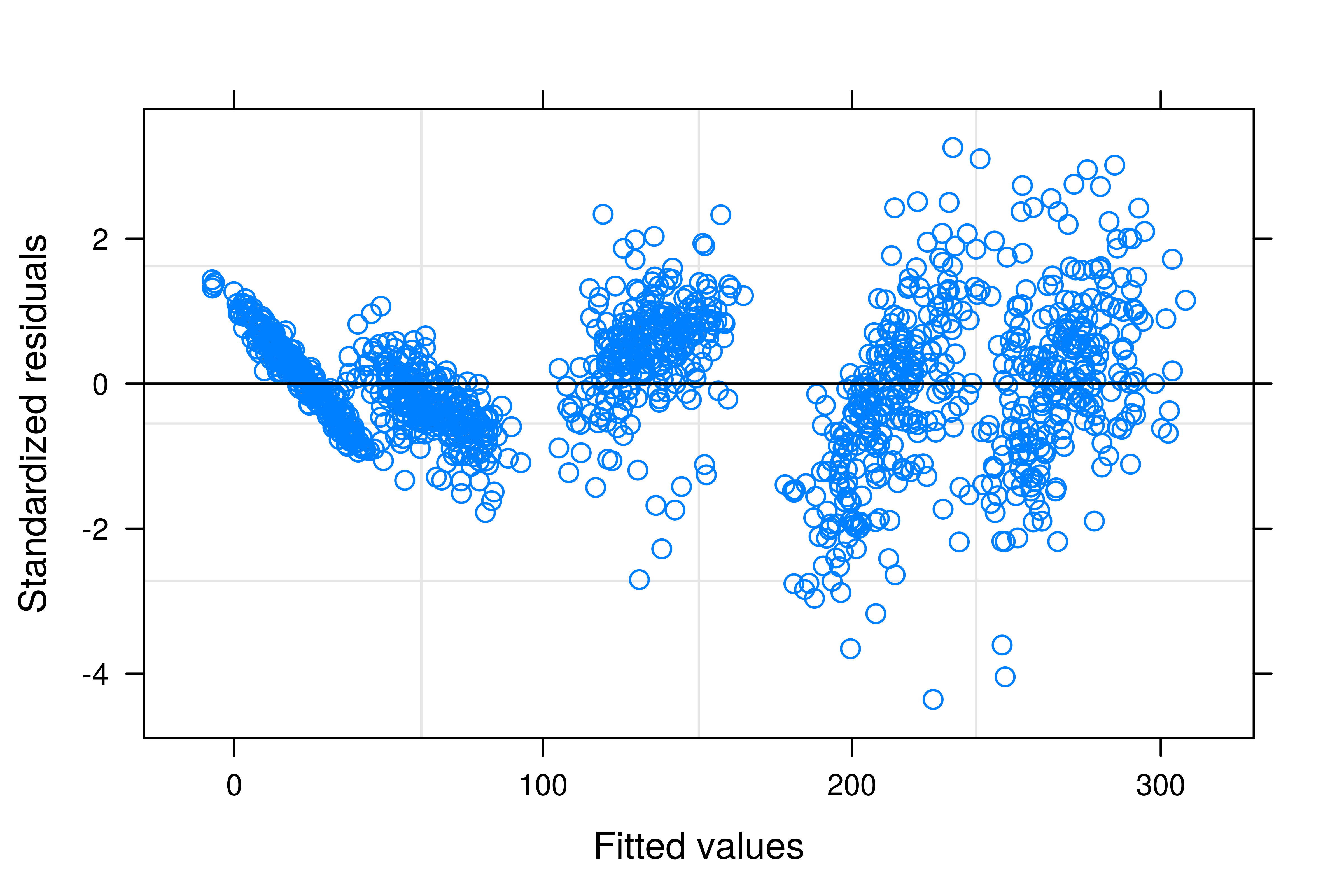}
	\caption{Residuals versus fitted values graph for the polynomial linear mixed model.}
	\label{res_clmm2}
\end{figure}

The correspondent graph of standardized residuals versus fitted values is given in Figure \ref{res_clmm2}. One of the issues appears to be diagnosed, that is, the residuals related to the height measurements taken at 45 days after seeding are more equally distributed around zero. However, there are still many residuals with high absolute values and the increase in the residuals' variability with time is evident.

Since the triple interaction of the linear effect of time with potassium silicate and soil tension levels is significant, one could think of testing the triple interactions of the factors with the quadratic and cubic effects of time as well. However, when these were included in the model neither of them was found to be significant using F-tests. Because the model with these two additional triple interactions is nested within the one with simple quadratic and cubic functions of time, another way of testing the contribution of the additional triple interactions is by means of a likelihood ratio test (LRT). The likelihood ratio test result between 
these two models shown in Table \ref{lrt} also favours the simpler one. The increase in the logarithm of the likelihood is too little to compensate for the number of degrees of freedom these terms require to be estimated.

\begin{table}[ht!]
	\centering
	\caption{Likelihood ratio test result of nested LMMs.}
	\label{lrt}
	\begin{tabular}{lllll}
		\hline
		Model           & Degrees of freedom & loglikelihood  & LRT Statistic  & p-value    \\ \hline
		M1 	 		    & 41     		   	& -5676.09 &				 &    \\  
		M2		 	    & 71   			    & -5660.39 & 31.40 			& 0.40    \\   \hline
	\end{tabular}
\end{table}

Therefore, the best model so far contains three nested levels of random effects (for plots, subplots and subsamples/plants), a fixed effect term of blocks, a triple interaction of soil tension and potassium silicate levels with the linear effect of time (and all correspondent two-level interactions and main effects) and the quadratic and cubic effects of time. This model does not account for the heterogeneity of variances though, as can be seen in Figure \ref{res_clmm2}. Even though there are three levels of random effects in the model, each of them accounts for the correlations among groups of observations defined by the physical arrangement of the experiment, which does not allow the model to estimate specific variances for groups of heights taken on the same day. Such a (more general) model can be obtained with a function to model the variance structure of the within-time errors using the time covariate. Following the notation of \citet{Davidian1995}, the general variance function for the multilevel mixed model (\ref{mod}) is expressed as

\begin{equation}
\mbox{Var}(\zeta_{ijklm}|\bepsilon_{ij},\bvarepsilon_{ijk},\bxi_{ijkl}) = \sigma^{2} g^{2}(\mu_{ijklm},\bv_{ijklm},\bdelta),
\end{equation}
where $\mu_{ijklm} = $ E$[y_{ijklm}|\bepsilon_{ij},\bvarepsilon_{ijk},\bxi_{ijkl}]$ (the expected plant height given the random effects or BLUP - best linear unbiased predictor), $\bv_{ijklm}$ is a vector of variance covariates, $\bdelta$ is the vector of variance parameters and $g(\cdot)$ is the variance function, assumed continuous in $\bdelta$.  The power variance function was used in this analysis because of its flexibility for modeling monotonic heteroscedasticity when the covariate used is bounded away from zero (which is the case with time since the first measurements were taken at 30 days). Using the time covariate, the power variance function is given by

\begin{equation}
\mbox{Var}(\zeta_{ijklm}) = \sigma^{2} \mbox{time}^{2\delta}_{ijklm}.
\end{equation}

\begin{figure}[!htb]
	\centering
	\includegraphics[width=10cm,height=7cm]{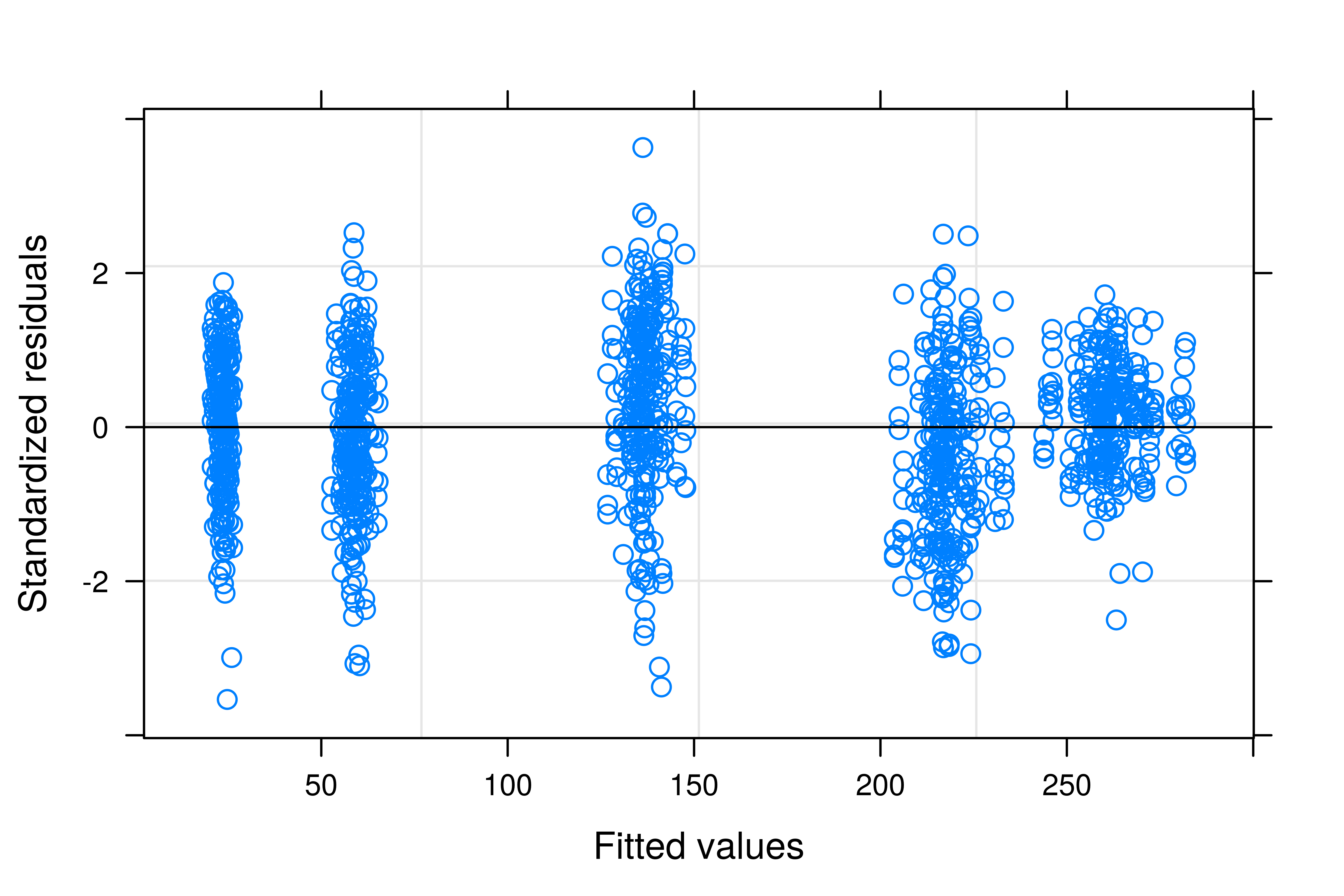}
	\caption{Graph of standardized residuals versus fitted values for the heteroscedastic mixed model.}
	\label{res_clmm3}
\end{figure}

Modifying the model with this variance function results in an heteroscedastic model that improves the fit over the previsous homoscedastic model for the plants' growth data. This can be seen by the likelihood ratio test result in Table \ref{lrt2}, which shows that there is a highly significant increase in the log-likelihood associated with the inclusion of the variance function. Furthermore, the residuals of the heteroscedastic model have now similar variances within days of observation, as shown in Figure \ref{res_clmm3}.
The estimated fixed effects changed very little by the use of the variance function.

\begin{table}[ht!]
	\centering
	\caption{LRT result of the heteroscedastic model over the homoscedastic model.}
	\label{lrt2}
	\begin{tabular}{lllll}
		\hline
		Model           & Degrees of freedom & loglikelihood  & LRT Statistic  & p-value    \\ \hline
		Homoscedastic   & 41     		   	& -5676.09 &				 &    \\  
		Heteroscedastic & 42   			    & -5268.10 & 815.97 			& \textless.0001    \\   \hline 
	\end{tabular}
\end{table}

\begin{figure}[!htb]
	\centering
	\includegraphics[width=13cm,height=10cm]{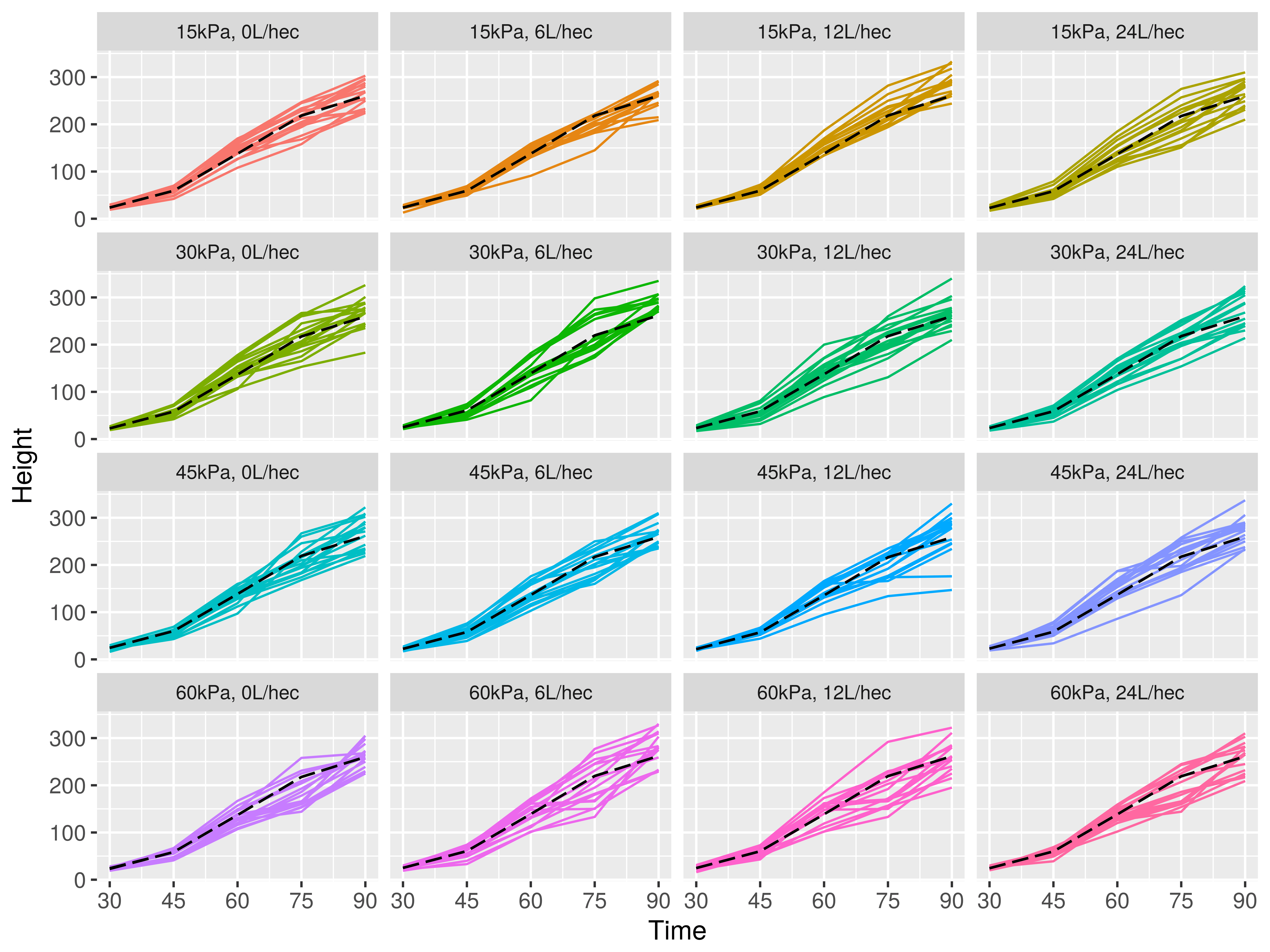}
	\caption{Observed and predicted heights with the final model.}
	\label{pred2}
\end{figure}

The plants' heights predicted by the heteroscedastic model along with the observed ones are depicted in Figure \ref{pred2} by treatment.
The traced lines denote the growths predicted by the model and the solid are the observed ones. The predicted growths look reasonable since for most treatments they are in the middle of the plant's growth lines observed. A striking feature in this graph is the lack of variation among the predicted lines, though. Given that the model includes random intercepts for plots, subplots and plants a higher distinction among the predicted lines' intercepts would be expected. 

\begin{figure}[!htb]
	\centering
	\includegraphics[width=7cm,height=10cm]{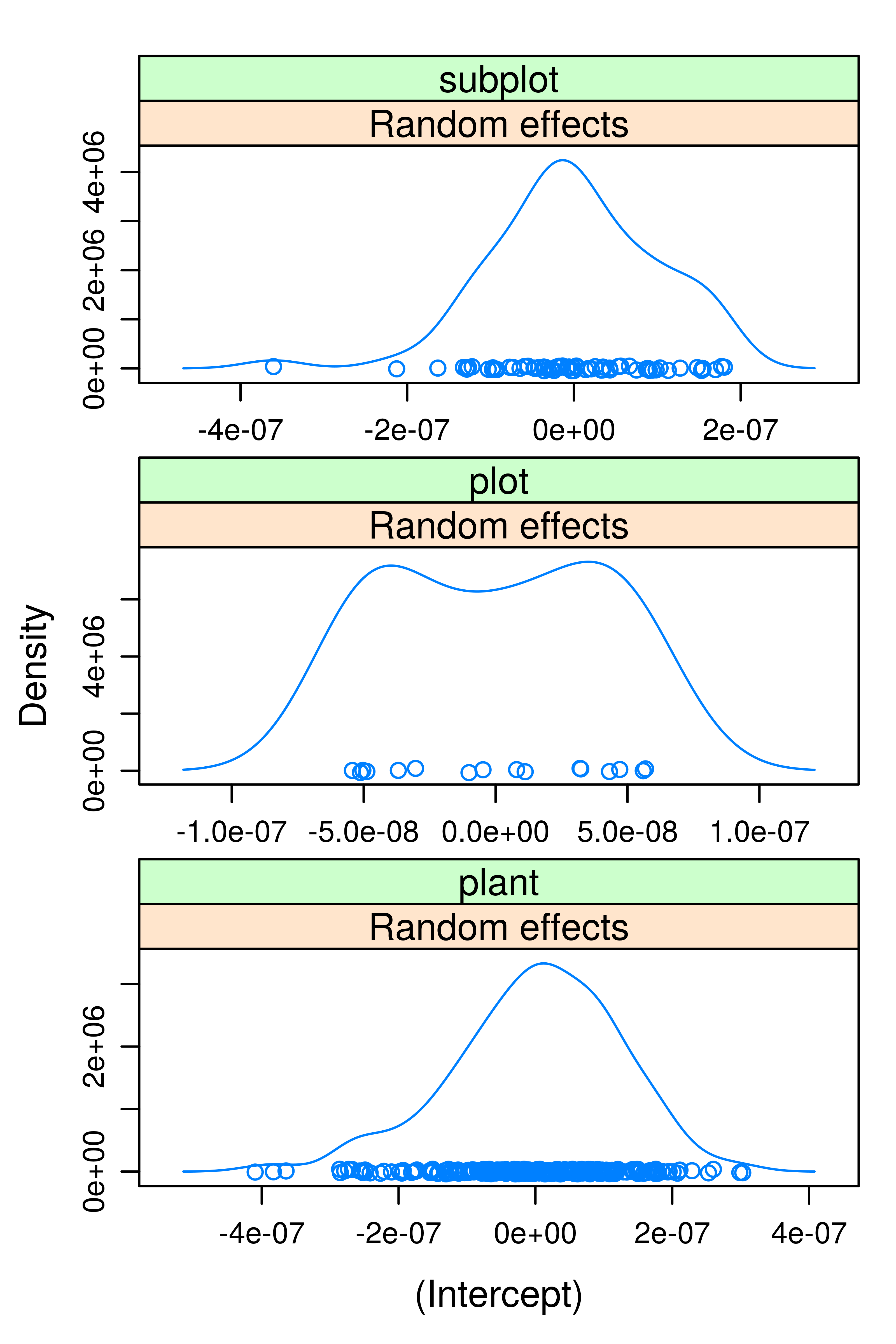}
	\caption{Random effects predicted by the final model.}
	\label{ranef}
\end{figure}

A visual evaluation of the random intercepts predicted by the final model is presented in Figure \ref{ranef}. There does not seem to be large deviations from the normality assumption by the graphs. However, the magnitude of the estimated intercepts is in the scale of tens of millions of units, that is, pretty much zero. This explains why the predicted plant growth lines in Figure \ref{pred2} are basically superimposed. Using likelihood ratio tests to test the contribution of one random effect at time using nested models all three gave significant contributions to the model, though. 

\subsection{Using function \textit{gamlss::gamlss()}}
\label{mgamlss}

The first implementation of the GAMLSS in the late 90's was written in GLIM4 (\citet{Francis1993}), which is nowadays almost an extinct software. The first \verb|R| implementation of the \verb|gamlss| function occured in the beggining of this century \citep{Akantziliotou2002}. In 2006 the first of a series of \verb|R| packages was released in CRAN (the Comprehensive \verb|R| Archive Network), the \verb|gamlss.dist|, that contained the initial distributions available to be used with the \verb|gamlss()| function implemented at that time. Currently, there are eleven \verb|R| packages of GAMLSSs available, each related to a specific aspect of this general type of regression models. This analysis was made with the base \verb|gamlss| package version 5.0-2 using probability distributions available in the \verb|gamlss.dist| package of the same version. 

A search on Google Scholar on 30th of September, 2017 for the word ``gamlss" retrieved 2.360 articles. GAMLSSs have already been applied to many scientific areas using parametric (linear and nonlinear) and semiparametric models that contain parametric and nonparametric smoothing terms, but no applications where GAMLSSs were used for mixed modeling purposes were found. The only related paper found was a comparison of GAMLSSs with hierarquical generalized models by \citet{Manco2011}, who found by simulations that both strategies showed similar results. 

Therefore, this analysis aims to illustrate how mixed models can be fitted using the GAMLSSs' methodology. The idea of comparing it with a classical and still largely used method is to highlight the advantages and disadvantages of this approach. Basically, a GAMLSS with random effects is a generalization of both linear and generalized linear mixed models by many reasons. The most important ones are that the exponential family assumption for the distribution of the response is relaxed and up to four parameters of the response distribution are possible to be modeled by functions of covariates, including random effects for as many of them as needed. The form of the normal mixed GAMLSS analogous to (\ref{modm}) is given by

\begin{equation}
\label{form}
\begin{array}{c}
\bY|\bepsilon,\bvarepsilon,\bxi \sim \mathbf{\mathcal{N}}(\bmu,\bsigma) \\

g(\bmu) = \bX_G\bmu + \bX_B\bbeta + \bX_A\balpha + \bX_L\blambda + \bX_{AL}\balpha\blambda + \bX_{T}\btau + \bZ_{SP}\bvarepsilon + \bZ_P\bepsilon + \bZ_{SS}\bxi \\

g(\bsigma) = \phi,
\end{array}
\end{equation}
where $\phi$ is the maximum likelihood estimate of the standard deviation of the plant's height in the scale of the link function $g(\cdot)$ used. 

The \verb|gamlss| library contains the functions \verb|gamlssNP()|, \verb|random()| and \verb|re()| for fitting random effects within a GAMLSS. However, only the last of them allows random effects to be estimated at different levels (multilevel modeling), which is required for the analysis of the sweet corn experiment. The fitting procedure of a mixed GAMLSS with \verb|re()| uses a local (internal to the GAMLSS fitting algorithm) normal approximation to the model's likelihood, known in the literature as penalized quasi likelihood (\citet{Breslow93}). The model's fitted values derive from the joint likelihood function of the response variable and the random effects' vectors, while inference is based on the conditional likelihood of the response given the random effects. Details of the fitting procedure can be found in chapters 3 and 10 of \citet{Stasinopoulos17}.

%
%

The form of defining a mixed GAMLSS in \verb|gamlss()| using \verb|re()| internally is exactly the same as with \verb|lme()|. This makes the fitting of mixed GAMLSSs with \verb|re()| straighforward to users of the \verb|nlme| package. Once the mixed GAMLSS is fitted and stored in an \verb|R| object, the function \verb|getSmo()| can be used to transform it to an \verb|lme| object, so that all methods used with mixed models fitted via \verb|lme()| become available, such as analysis of variance via the \verb|anova()| function and random effects estimates by level via \verb|ranef()|. The only tool of \verb|lme()| that is not available via \verb|re()| is to define variance functions for the within group heteroscedastic structure, which was used in the previous analysis. This is because the philosophy of GAMLSSs is to handle problems at the observational level (such as overdispersion and heterogeneity of variances) using more general distributions for the response variable instead of trying to overcome them with less flexible ones.

Apart from having the facilities of \verb|lme()| through the intercace with \verb|re()|, the \verb|gamlss()| function also provides its own functionalities for fitting and assessing regression models. Among the functions that were useful in this analysis were the \verb|wp()| and \verb|plot()| functions. The \verb|wp()| function provides single or multiple worm plots for \verb|gamlss| fitted objects. Worm plots are detrended QQ-plots introduced by \citet{vanBuuren01} in order to find intervals of an explanatory variable where the model does not adequately fit the data. The well known \verb|plot()| function of \verb|R| has a personalized version in the \verb|gamlss| library. For a \verb|gamlss| fitted object, it produces a two by two matrix of plots, that contains a plot of the residuals  against fitted values of the $\mu$ parameter, against an index of specific covariate, a kernel density estimate of the residuals and a QQ-normal plot of the residuals. It is important to state that GAMLSSs use normalized quantile residuals of \citet{Dunn96} (henceforth referred to simply as residuals), which are more general than other types of residuals in the sense that they always have a standard normal distribution when the assumed model is correct, whatever the distribution of the response variable. 

\begin{figure}[!htb]
	\centering
	\includegraphics[width=13cm,height=8cm]{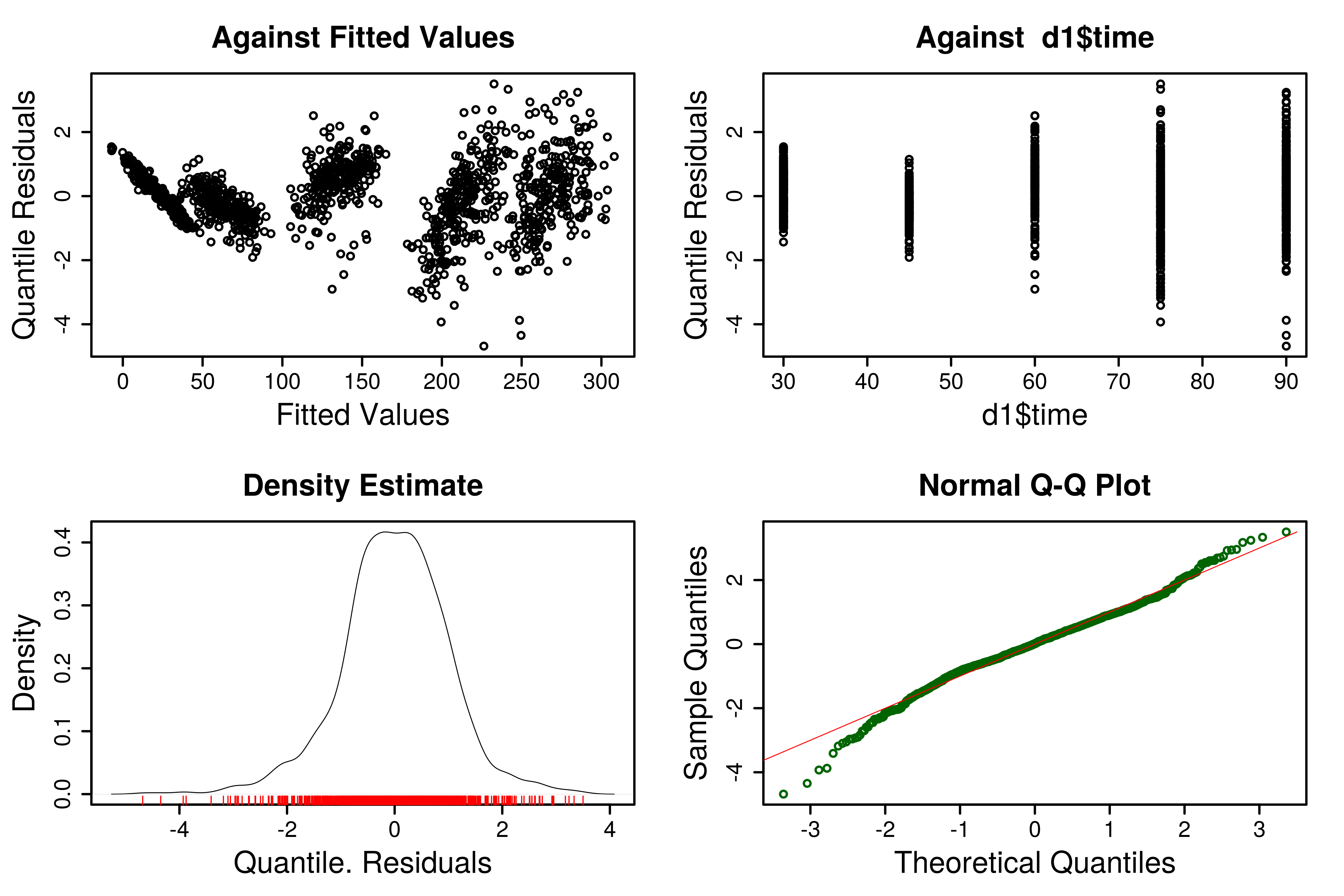}
	\caption{Plots of residuals for mixed model with NO distribution.}
	\label{plot_no}
\end{figure}

\begin{figure}[!htb]
	\centering
	\includegraphics[width=10cm,height=7cm]{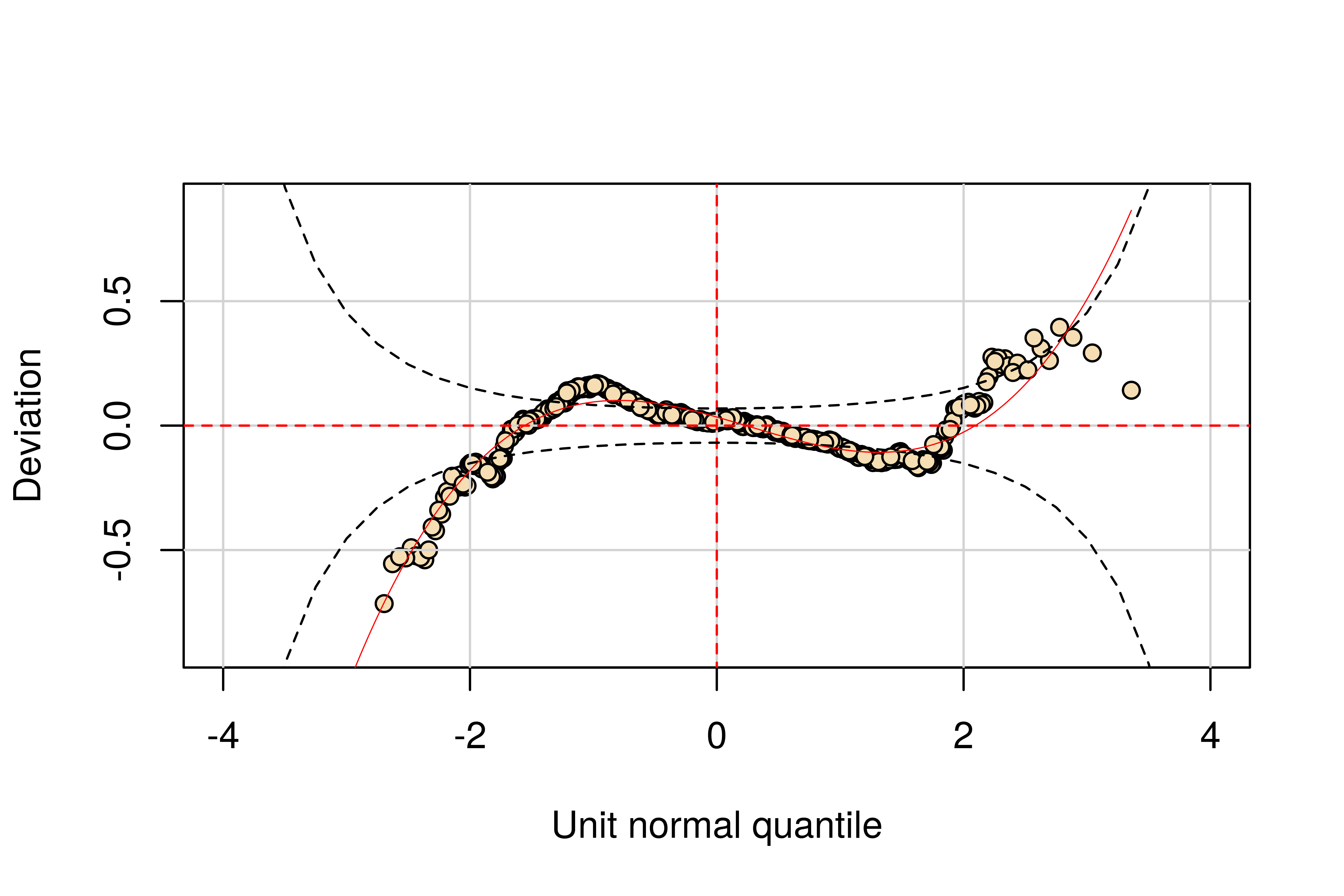}
	\caption{Worm plot of residuals for mixed model with NO distribution.}
	\label{wp_no}
\end{figure}

Initially, a mixed model with the same random and fixed effects terms used in the final model in the previous analysis was fitted, also assuming normal distribution (NO) for the conditional response variable. F-tests' results for selected fixed effects terms in this model are shown in Table \ref{anova.no}. They are exactly the same as Table \ref{lmm2}. A graph of the residuals for this model obtained with the personalized \verb|plot()| function is shown in Figure \ref{plot_no}. The top left plot in this figure is equal to Figure \ref{res_clmm2}. The plot on the right of it shows the residuals against the days when the plant's heights were taken, highlighting the already known heterogeneity of variances by time. The two bottom plots show that the normality assumption does not seem too bad for this model. However, the worm plot presented in Figure \ref{wp_no} is a little more enlightening about it. If the normality assumption is not to be rejected, at least 95\% of the quantile residuals should stay between the two elliptic curves (they are approximate point-wise 95\% confidence intervals for the residuals), which is clearly not the case. Another feature of worm plots is that different shapes indicate different inadequacies in the model. The worm plot in Figure \ref{wp_no} has an S-shape form with left bent down, which indicates that the tails of the fitted distribution are too light (the meanings of all worm plots' shapes  are given in section 12.4 of \citet{Stasinopoulos17}). 

\begin{table}[ht!]
	\centering
	\caption{F-tests for selected terms of the mixed GAMLSS with normal distribution.}
	\label{anova.no}
	\begin{tabular}{lllll}
		\hline
		Term           & Degrees of freedom & F-value    & p-value    \\ \hline
		time:tension:silicate  & 9     & 3.75      & 0.0001         \\
		time$^2$       	   & 1     & 69.15      & \textless.0001         \\
		time$^3$ 	   & 1     & 180.52     & \textless.0001    \\   \hline
	\end{tabular}
\end{table}

\begin{figure}[!htb]
	\centering
	\includegraphics[width=13cm,height=8cm]{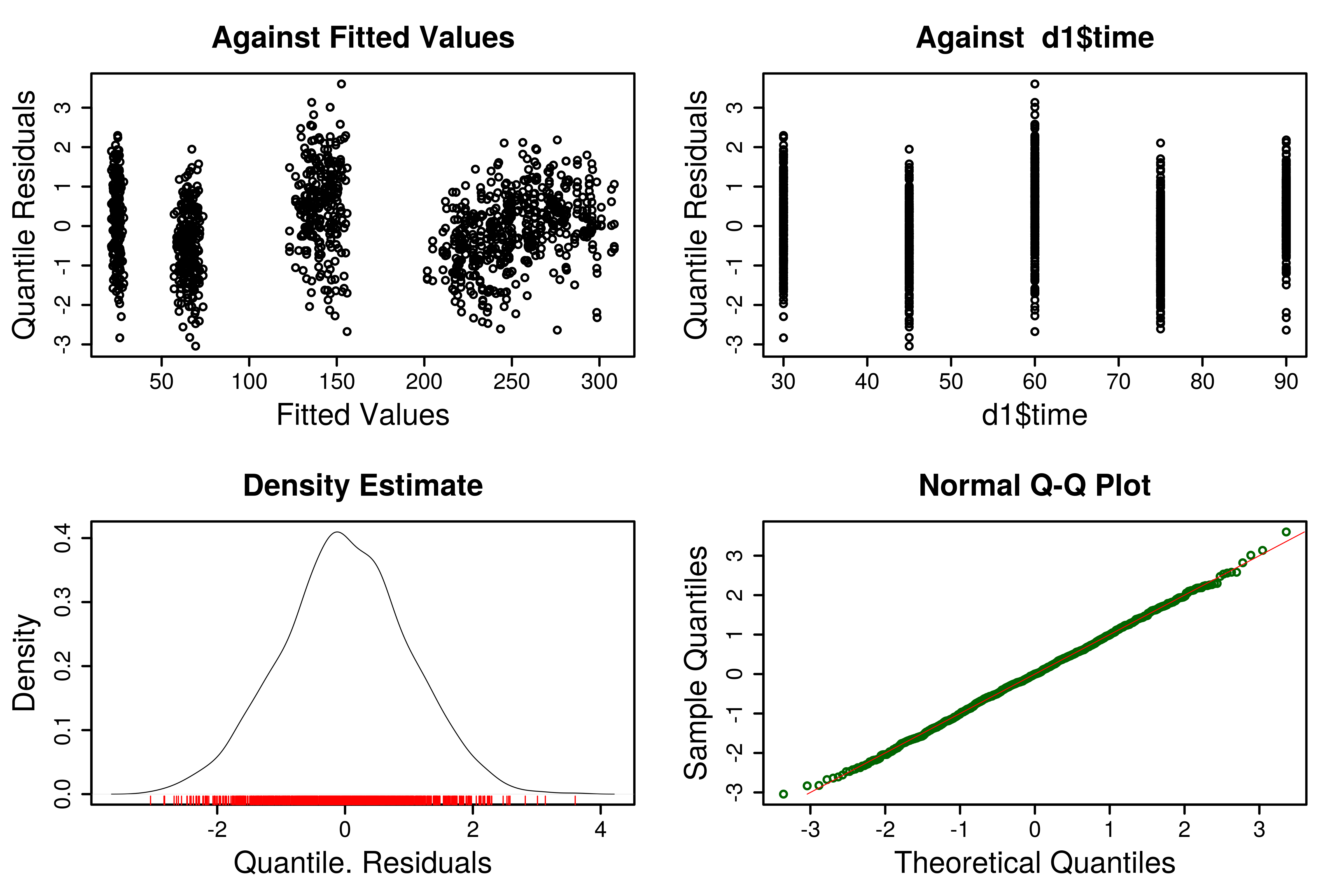}
	\caption{Plots of residuals for mixed model with GG distribution.}
	\label{plot_gg}
\end{figure}

\begin{figure}[!htb]
	\centering
	\includegraphics[width=10cm,height=7cm]{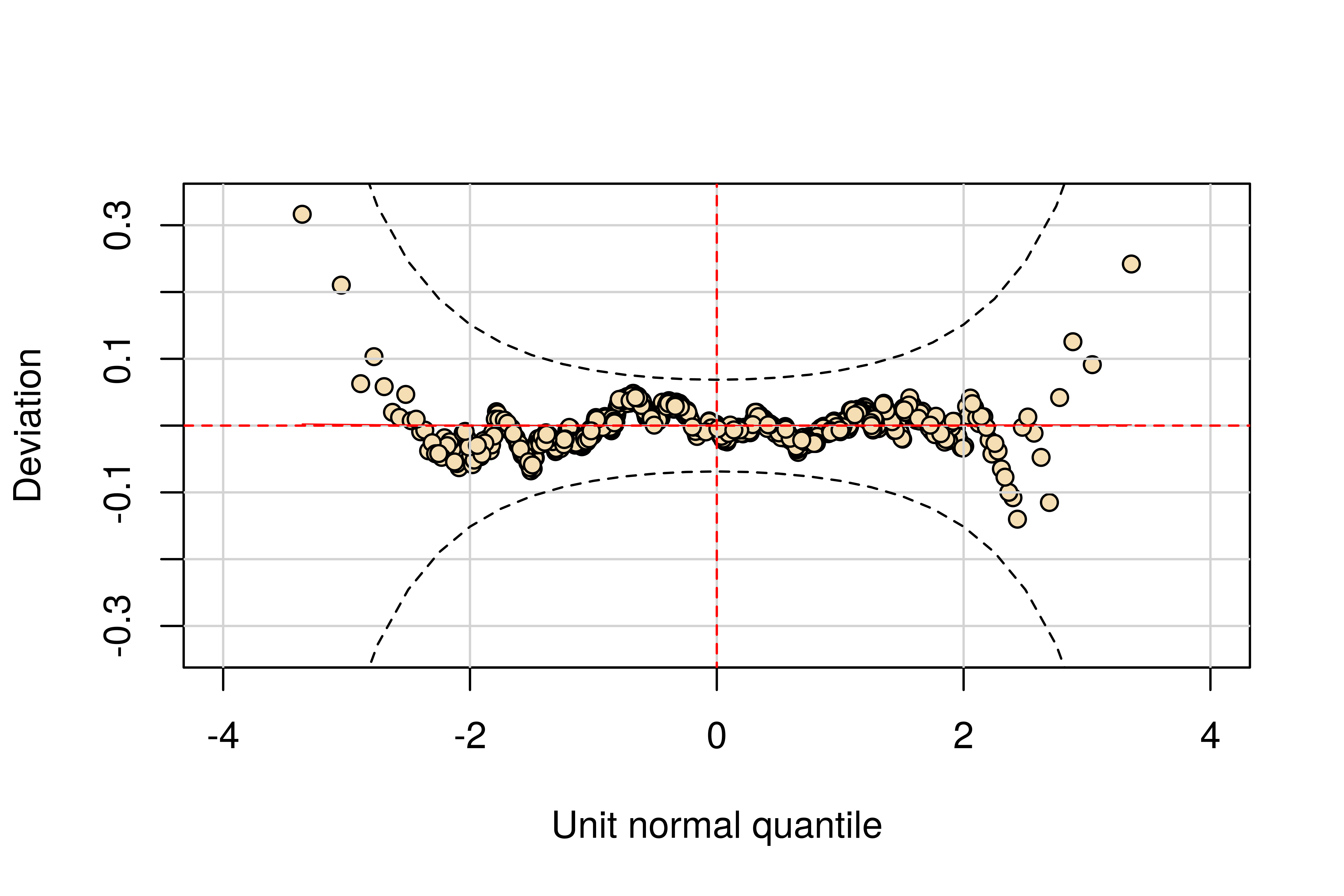}
	\caption{Plots of residuals for mixed model with GG distribution.}
	\label{wp1_gg}
\end{figure}

A more general alternative distribution that allows for heavier tails than the normal is the generalized gamma (GG) (Stacy (1962)\footnote{Stacy, E. W. (1962). A generalization of the gamma distribution. The Annals of mathematical statistics, 1187-1192.}). The GG is a three-parameter distribution with mean equal to the first parameter $\mu$ and variance equal to $\mu^2\sigma^2$, a function of the first two parameters ($\sigma$ is the second parameter). The third parameter $\nu$ is related to the skewness of the distribution. The default link functions for each parameter were used when modeling, which are the log function for $\mu$ and $\sigma$ and the identity for $\nu$. A mixed GAMLSS was fitted using the GG distribution with the same model formula for fixed and random effects as before. Using F-tests the results found were very similar to the previously obtained ones with the normal distribution. This reinforces that there is a triple interaction effect of the linear function of time with potassium silicate and soil tension levels and a cubic polynomial function of time is needed. The model also estimated the $\sigma$ and $\nu$ parameters as significantly different from zero. The two by two graph of the residuals for this model is presented in Figure \ref{plot_gg}. From the residuals vs fitted values plot can be seen that the heterogeneity of variances has been addressed by the use of the GG distribution. The plot of residuals vs time confirms that the residuals' variability is similar for all days of measurements. Both bottom plots agree in that there is no deviation from the normality assumption for this model overall. The worm plot for this model (Figure \ref{wp1_gg}) confirms that by displaying the solid fitted curve of the residual points basically on the x-axis and with all points inside the two elliptic curves.

\begin{figure}[!htb]
	\centering
	\includegraphics[width=10cm,height=7cm]{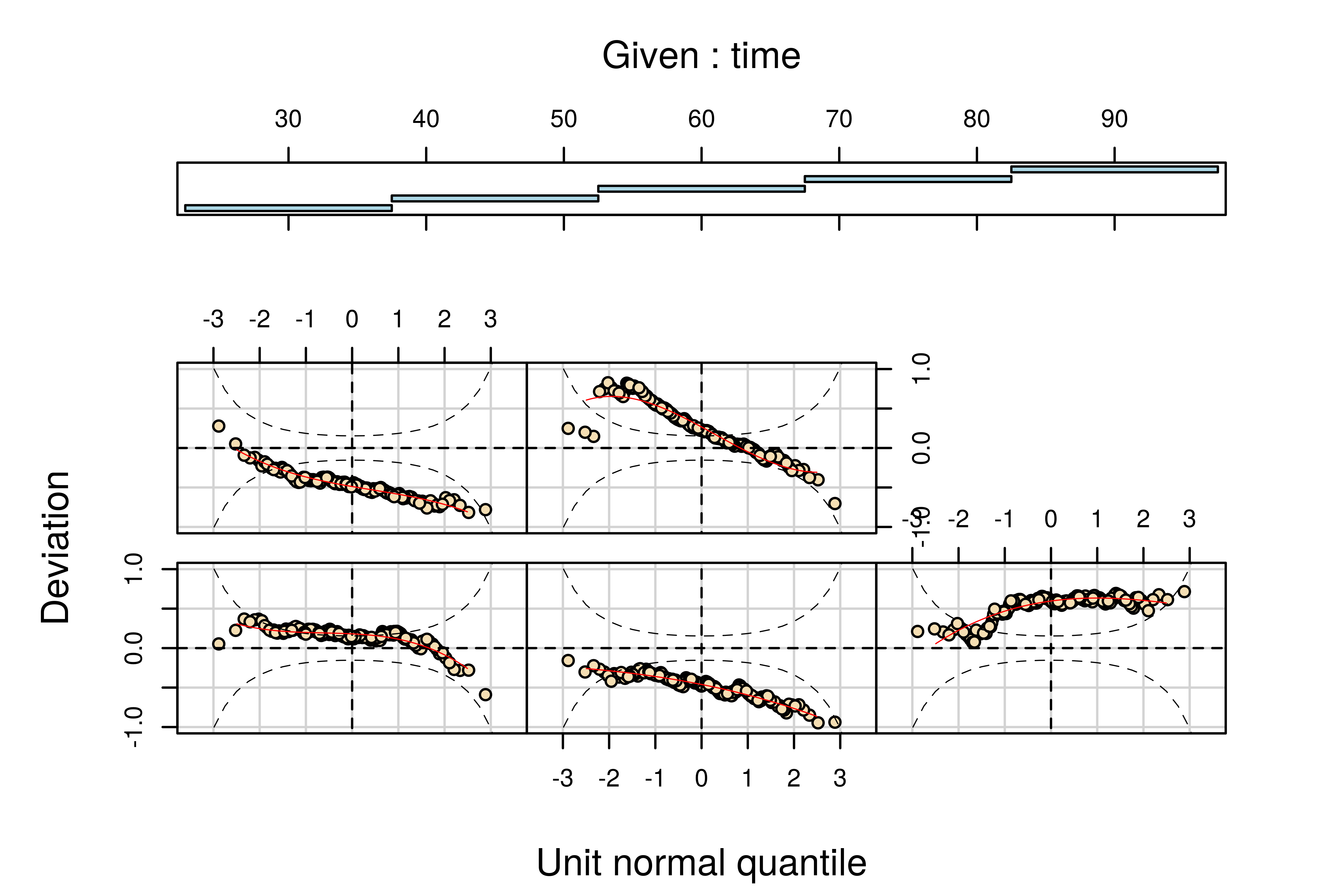}
	\caption{Worm plots by time for mixed model with GG distribution.}
	\label{wp_gg}
\end{figure}

This overall worm plot is equal to the normal QQ-plot in Figure \ref{wp_gg} detrended by subtracting the line. It highlights deviations of the observed residuals to their approximate expected values represented by the horizontal dotted line. However, there might be failures in the model within ranges of the explanatory variables yet. These are minor inadequacies in the model but their diagnose is still recommended, specially for the most important explanatory variables in the analysis (\citet{Stasinopoulos17}, page 428). This is possible to be check for GAMLSSs with multiple worm plots by intervals of the explanatory variables (or by levels of factors). When applied to the levels of soil tension and potassium silicate factors in the mixed GG model no misfits were found. For the time covariate five worm plots were generated with non-overlapping intervals of equal number of observations and are displayed in Figure \ref{wp_gg}. The upper panel gives the interval ranges (each day of observation in this case) and the worm plots below are read along rows from bottom left to top right, made with the residuals on each interval in order. For example, the bottom right worm plot was made with residuals of heights taken at 60 days (the third time interval). Contrasting with the previous worm plot shown, in this figure all worm plots appear to have problems. Each worm plot contains a thin solid line that is a cubic fit on the residuals. When fitting multiple worm plots with \verb|wp()| it is possible to get the coefficients of these cubic fits, which help understanding the inadequacies in the worm plots. The estimated coefficients (intercept $\hat{\beta_0}$, linear $\hat{\beta_1}$, quadratic $\hat{\beta_2}$ and cubic $\hat{\beta_3}$) of the cubic fits in these worm plots are given in Table \ref{cf}.

\begin{figure}[!htb]
	\centering
	\includegraphics[width=10cm,height=7cm]{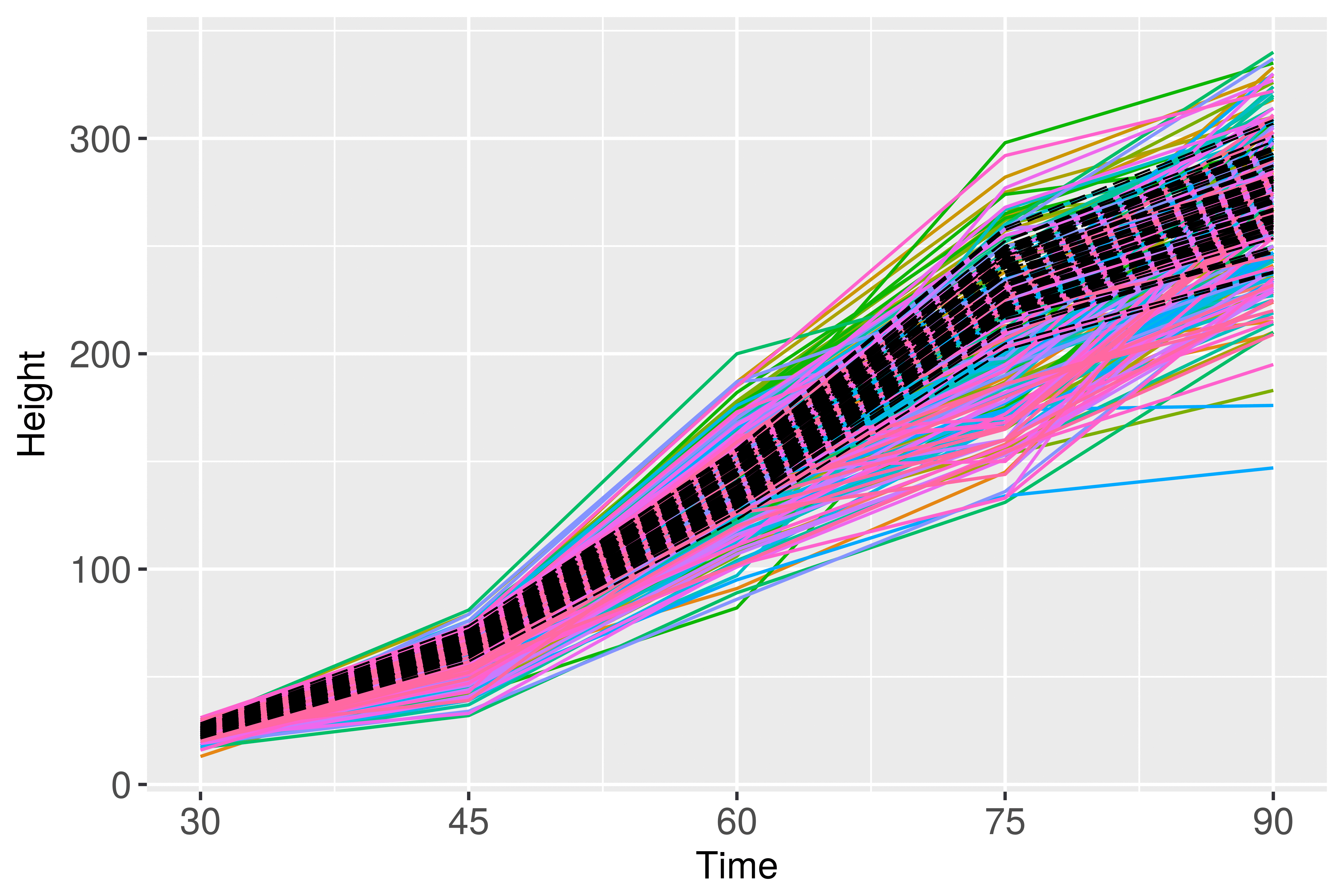}
	\caption{Plants' heights observed and predicted with GG mixed model.}
	\label{predg}
\end{figure}

\citet{vanBuuren01} classify the absolute values of these cubic fits' coefficients. For the intercept and the linear coefficient they established that estimates above 0.1 are considered misfits that (according to \citet{Benard1953} and \citet{VanZwet1964}) indicate high differences in the mean and variation of the theoretical model residuals and the fitted residuals. Similarly, they considered threshold values for quadratic and cubic absolute estimates as 0.05 and 0.03 and argue that they suggest inadequacies in the residuals' skewness and kurtosis, respectively. Applying these rules to the values on Table \ref{cf} it can be concluded that there are inadequacies in the mean of the residuals at all times (since the first column has all estimates in absolute values above 0.1), in the variance for some days and that there are not misfits in the skewness and kurtosis of the residuals. This residual evaluation can also be related directly to the height predictions obtained with this model, shown in Figure \ref{predg} in black dotted lines over the observed ones. For example, the highly negative intercept for the cubic fit at 45 days in Table \ref{cf} means that the mean of the fitted distribution at that time was too high compared to the observed one, generating negative residuals in the bottom middle worm plot. This translates to superestimated plants' heights on average at that time, as can be seen in Figure \ref{predg}.

\begin{table}[]
	\centering
	\caption{Coefficients of cubic fits in worm plots of GG mixed model.}
	\label{cf}
	\begin{tabular}{ccccc} 
		\hline
		Time  & $\hat{\beta_0} $  & $\hat{\beta_1}$ & $\hat{\beta_2}$ & $\hat{\beta_3}$  \\ \hline
		30 days & 0.1806  & -0.0291 & -0.0251 & -0.0135  \\
		45 days & -0.4605 & -0.1162 & -0.0163 & -0.0008  \\
		60 days & 0.5964  & 0.0790  & -0.0454 & 0.0039   \\
		75 days & -0.4880 & -0.1002 & 0.0116  & -0.0092  \\
		90 days & 0.2662  & -0.3170   & -0.0187 & 0.0215  \\ \hline
	\end{tabular}
\end{table}

\citet{Stasinopoulos17} comment that in general it will not be possible to build a model without areas of misfits for all the covariates included. Even though very sensitive, this detailed residual analysis gives valuable information about the model fitting according to ranges of the explanatory variables. Furthermore, it shows where specific failures in the model are and what needs to be done to minimize them. In this analysis, the problem with the mean has to do with the flexibility of the chosen function for the time effect. Probably a nonlinear or a smoothing function of time would account better for the unequal growth rates. However, since there are no problems with the residual analysis by neither of the factors, it is unlikely that the significant interaction effect would change by modifying the time function. 

\begin{figure}[!htb]
	\centering
	\includegraphics[width=7cm,height=10cm]{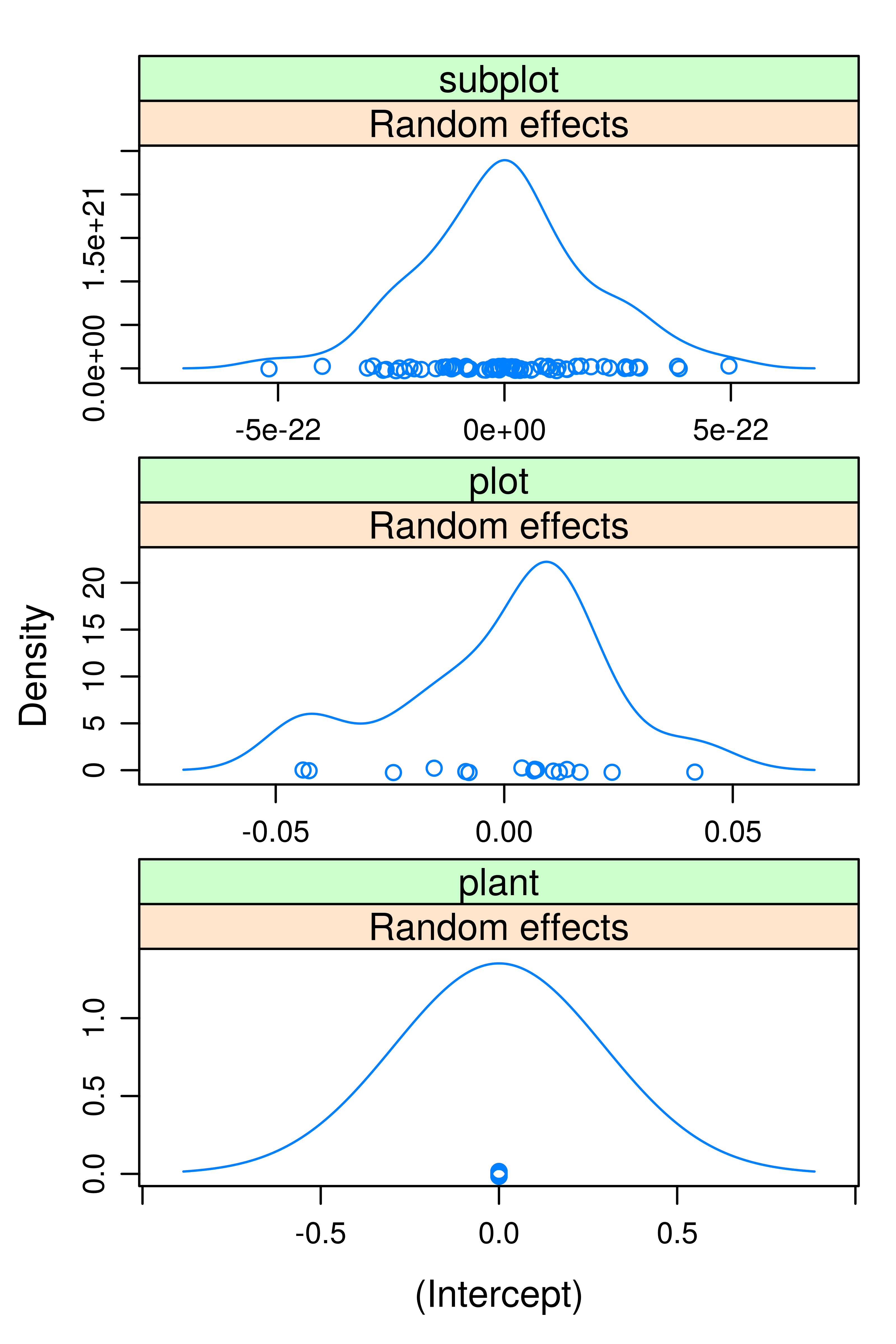}
	\caption{Estimated random effects with the GG mixed model by level.}
	\label{ranef2}
\end{figure}

Finally, the predicted random effects for the GG mixed GAMLSS are shown in Figure \ref{ranef2}. Evaluations of the normality assumption might be unreliable due to the number of observations they take into account, but overall there does not seem to be issues with normality. Differently than in the previous analysis, here the magnitude of the random effects vary considerably by level and are much higher in the plots and plants levels than the correspondent ones obtained by the linear mixed model. This is also noticeable from the variability of the height predictions in Figures \ref{pred2} and \ref{predg}. 

\section{Conclusions}

In this paper a comparison of two methodologies for analysing a multilevel mixed model was described. In the first, that used the \verb|nlme::lme()| function, the heterogeneity of variances was handled by defining a power variance function for the residuals by the time covariate. The final model contained three levels of random effects whose values were predicted in the scale of $1 \times 10^{-7}$, resulting in similar predictions for plants that received the same treatment. The significant effects obtained with this model were a triple interaction of soil tension and potassium silicate factors with the linear effect of time, apart from significant effects of the quadratic and cubic functions of time. The second analysis used the GAMLSSs methodology with random effects to model the plant growths. The initial model fitted with \verb|lme()| assuming normality for the residuals was discarded and a more flexible one with the generalized gamma distribution was used instead.  The estimated random effects varied in magnitude by levels which resulted in more variable predictions of the plants' growths. The GG mixed GAMLSS suggested the same fixed effects terms as the linear mixed model. 

The approaches used to deal with inadequacies in the model were different in the two analysis. The heteroscedasticity was accommodated in the second analysis using a more flexible distribution for the response variable, while in the linear mixed model a more general form for the variance matrix of the residuals was defined. However, in the linear mixed model a detailed residual analysis using worm plots by time was not made. In fact, checking the normality assumption with a shapiro-wilk test for the final mixed model resulted in a p-value in the order of $1 \times 10^{-8}$, while for the mixed GAMLSS with GG distribution the same test gave $0.91$ as the descriptive level. The AIC criterion for the LMM is 10620,21 and for the GG GAMLSS is 10425,14, also favouring the second. Furthermore, the \verb|re()| interface of \verb|gamlss()| allows the same facilities of \verb|lme()| as well as a wide range of response distributions to be used and the possibility to model other parameters than the mean explicitly with covariates.

The result of the experiment was not dependent on the method of analysis used, though. The conclusion is there are significantly different growth trends in plants that were on soil with different tension levels and received different doses of potassium silicate. However, which treatment combinations differ among themselves and why would have to be further explored, since the graph of average growth lines per treatment in Figure \ref{height.treat} is not clear about it. Furthermore, another modeling approach should be investigated to reduce inadequacies in the model with respect to the time covariate. Possible alternatives could be to use a logistic function modeling this functions' parameters with random and fixed effects or to smooth the time effect with a nonparametric function such as a P-spline.

\clearpage

\renewcommand\bibname{References} 
\bibliographystyle{./referencias/genetics} 

\clearpage

\end{document}